\documentstyle[11pt,aaspp4]{article}
\slugcomment{To appear in the Astronomical Journal}
\lefthead{Gioia et al.}
\righthead{RXJ1716.6+6708}
\def\deg{\hbox{$^\circ$}}
\def\sun{\hbox{$\odot$~}}
\def\arcmin{\hbox{$^\prime$}}

\def\etal{{\it et al.} }
\def\hawaii{Hawai$'$i~}
\def\arcsec{\ifmmode^{\prime\prime}\;\else$^{\prime\prime}\;$\fi}
\def\rx{RXJ\thinspace 1716.6$+$6708}

\begin{document}

\title{RXJ\thinspace 1716.6+6708: a young cluster at z$=$0.81}

\author{I. M. Gioia\altaffilmark{1,2}, J. P. Henry\altaffilmark{2},
C. R. Mullis\altaffilmark{2}, H. Ebeling}
\affil{Institute for Astronomy, University of Hawaii, 2680 Woodlawn 
Drive, Honolulu, HI 96822}

\and

\author{A. Wolter}
\affil{Osservatorio Astronomico di Brera, Via Brera 28, I-20121, Milano, Italy}

\altaffiltext{1}{Also Istituto di Radioastronomia del CNR,
Via Gobetti 101, I-40129, Bologna, ITALY}

\altaffiltext{2}{Visiting Astronomer at Canada-France-Hawaii Telescope,
operated by the National Research Council of Canada, le Centre
National de la Recherche Scientifique de France and the University
of Hawai$'$i, and at the W. M. Keck Observatory, jointly
operated by the  California Institute of Technology, the
University of California and the National Aeronautics and Space
Administration.}

\begin{abstract}
		
Clusters of galaxies at redshifts nearing one are of special
importance  since they may be caught at the epoch of
formation. At these high redshifts there are very few known clusters. 
We present follow-up ASCA, ROSAT HRI and Keck LRIS observations 
of the cluster 
\rx ~which  was discovered during the optical identification of X-ray 
sources in the North Ecliptic Pole region of the ROSAT All-Sky Survey.
At z$=$0.809, \rx ~is the second most distant X-ray selected cluster
so far published and the only one with a large number of
spectroscopically determined cluster member velocities. 
The optical morphology of \rx ~resembles an inverted S-shape
filament with the X-rays coming 
from the midpoint of the filament. The X-ray contours have an
elongated shape that roughly coincides with the weak lensing
contours. The cluster has a low temperature,
$kT=5.66^{+1.37}_{-0.58}$ keV, and a very high velocity dispersion
$\sigma_{los}=$1522$^{+215}_{-150}$ km s$^{-1}$. While the 
temperature is commensurate with its X-ray luminosity  of
(8.19$\pm$0.43)$\times10^{44}$ h$_{50}^{-2}$ erg s$^{-1}$ (2$-$10 keV
rest frame), its velocity dispersion is much higher than expected 
from the $\sigma-T_X$ relationship  of present-day clusters with 
comparable X-ray luminosity.  \rx ~could be an example of a
protocluster, where matter is flowing along filaments and the X-ray
flux is maximum at the impact point of the colliding streams of matter.

\end{abstract}


\section{Introduction}

In hierarchical theories of structure formation, clusters of galaxies form
from the high peaks in the original density field, thus they
provide crucial constraints on the shape, amplitude, and
temporal evolution of the primordial mass fluctuation spectrum.
Despite their importance, the statistics for the abundance of high-z ($>$0.5)
clusters are poor since they are so difficult to locate. Optical surveys
of distant galaxy clusters are well known to have serious statistical
shortcomings such as effects of superpositions of unvirialized systems
(see among others \cite{fre90}; \cite{har97}). In addition, 
at high redshifts it becomes  difficult to detect
enhancements in the galaxy surface density against the overwhelming 
field galaxy population (see \cite{oke98} for the success rate
of optical high-z cluster searches). 

One of the cleanest ways to avoid  sample contamination 
is  the selection of high-z clusters  by means 
of their X-ray emission. X-ray surveys are sensitive enough
to detect distant clusters. Examples include MS1137$+$66, 
at z$=$0.78 and MS1054$-$03 at z$=$0.83 in the Medium Survey
(\cite{gio90} and \cite{gio94}; see also the detailed study 
on MS1054$-$03 by \cite{don98a}).
Other X-ray surveys being conducted with ROSAT archive data 
are also  finding distant clusters, for example the WARPS
cluster at 0.83 (RXJ0152.7-1357, Ebeling et al. 1998) and the
clusters discovered in the RDCS survey (\cite{ros95}; 
\cite{ros98}). The RDCS
holds the record for the most distant X-ray selected 
cluster  known at z$\sim$1.26 (Rosati et al. 1999).
These X-ray surveys are demonstrating that they can go as 
deep as optical searches in locating distant clusters. 

Other techniques are successfully isolating clusters.
Clustering around radio galaxies at redshift equal or greater than 1 
has been reported  (see among others 3C 184,
z$=$0.996, \cite{del97}; MRC0316$-$257, z$=$3.14, \cite{olf96}; 3C
294, z$=$1.786, \cite{cra96}; 3C 324, z$=$1.206,
 \cite{sma95}). A new promising technique which is currently providing 
interesting results is the search for distant clusters in the near-IR
(ClG J0848$+$4453, z$=$1.27 from 8 cluster members, \cite{sta97}).
However, with the exception of the survey of nine high redshift 
optically-selected clusters by Oke and collaborators
(\cite{oke98}, \cite{lub98}, \cite{pos98}), most of these distant 
structures have spectroscopic determinations for less than one dozen 
cluster members, at best. 

Our group at the Institute for Astronomy in \hawaii
has been involved for several years in the optical
identification of all the sources found in  84.7 deg$^{2}$ 
centered at the North Ecliptic Pole (NEP) of the  ROSAT 
All-Sky Survey (RASS, \cite{vog96}). The NEP region is
the deepest area of the RASS. We reported in  Henry et al. (1997)  
the detection of a distant cluster, \rx, at z$=$0.8
with only 33 net photons. This weak detection 
revealed a very interesting object. The spectroscopy in 
hand at the time of the Henry et al. (1997) paper was limited to only 12 
cluster members but there was already an indication  (see their
Fig. 1) that the galaxies in the inverted S-shaped
filament were all part of the X-ray cluster. We report here
follow-up observations of \rx ~at optical and X-ray wavebands.

Very high-redshift X-ray selected clusters,
such as the one presented here, are important to study. 
Optically selected clusters tend to be under-luminous 
X-ray sources (\cite{bow97} and reference therein),
an indication that generally they are  not massive objects. 
As more accurate observing capabilities  become available,
the number of distant known clusters will steadily increase,
and the counts may severely constrain cosmological models.
However, we are still far from having large complete samples of
high-z clusters to test the predictions of cosmic structure formation
theories at these redshifts.

Throughout this paper, we assume H$_{0}=50$
km s$^{-1}$ Mpc$^{-1}$, and q$_{0}=0.5$,
unless otherwise stated. At the redshift of the cluster, 
the luminosity distance is 5.56 h$_{50}^{-1}$ Gpc, the 
angular size distance is 1.70 h$_{50}^{-1}$ Gpc,
and the scale is 8.24 h$_{50}^{-1}$ kpc per arcsec.

\section{Observations}

The discovery data of \rx ~are reported in  Henry et al. (1997).
Here we present the optical spectroscopy performed at Keck-II and
the X-ray follow-up observations
acquired with the HRI on ROSAT and with the ASCA satellite.
We first describe the optical spectroscopy which gives a very 
high value for the velocity dispersion. Then we present the
ASCA spectra which show that \rx ~has
a temperature commensurate with its bolometric luminosity
but not with its velocity dispersion.
Finally, we describe in detail the HRI data and present an X-ray
image revealing a morphology which is elongated in the same
direction as the optical galaxies.

\subsection{KECK slit-mask Spectroscopy}

The region of \rx ~was observed spectroscopically  on July 10 and 11,
1997, with the Low Resolution and Imaging Spectrograph (LRIS, \cite{oke95})
and  slit-masks on the Keck-II telescope.  The 300 l/mm grating 
blazed at 5000 \AA\ combined with our grating angle yielded
a wavelength coverage of approximately  
4500 \AA\ to 9500 \AA, and a pixel size of 2.45 \AA.
A GG495 filter was used to eliminate the overlapping second order
spectrum so as not to have any contamination blueward of
9500 \AA.  Given the fact that galaxies are extended objects even at this
redshift, we have used a slit of 1.5\arcsec in width, which gives
a reduced spectral resolution of 16 \AA. For the selection of
the objects to observe, we have used a very  deep 3 color (B, R and I)
image kindly supplied by Doug Clowe  and published in 
Clowe et al. (1998). The UCSCLRIS package designed by Drew
Phillips and collaborators at Lick Observatory
was used to prepare the slit-mask files.  We designed
four masks at different position angles in order to cover 
the central  part of the cluster. The four masks were exposed for 
10,350s, 10,800s, 13,500s and 10,300s respectively. After each mask,
exposure flat fields and arc calibrations were also obtained.
The data were reduced using the standard IRAF package routines for
2-D spectra images.  

Some of the objects were observed more than once
in different masks. In the end, out of 62 unique objects observed,
37 proved to be cluster member galaxies. The galaxies in the cluster are
marked in Figure~\ref{image} and listed in Table 1. For each galaxy the 
(J2000) coordinates, measured velocity, 1$\sigma$ error, and
redshift are given. In the last column the main spectral features are
also noted. The velocity histogram for the 37 member galaxies 
is shown in Figure~\ref{vel_hist}. Using the Timothy Beers' program, ROSTAT, 
we obtain an average redshift and velocity dispersion (corrected to first
order for cosmological redshift by dividing by 1$+$z)  which are
z$=$0.8090$\pm$0.0051 and $\sigma_{los}=$1522$^{+215}_{-150}$  km s$^{-1}$.
Using different methods (\cite{bee90}, \cite{dan80}) we obtain the
same velocity dispersion within the uncertainties. This high velocity 
dispersion   was already noted by  Henry et al. 1997 with only 12 
galaxy redshifts (the 37 cluster members presented here include the 12
cluster galaxies of Henry et al. 1997).
The old and new results are consistent  with each other within the 
uncertainties. We will discuss the implications of the high 
velocity dispersion for this cluster in Section 3.

\subsection{ASCA Spectroscopy}

We observed \rx ~with ASCA (\cite{tan94}) on June 9 and 11,
1996 for 115,779 and 53,191 seconds respectively.  
The ASCA instrumentation consists of four independent detectors, two
Solid-State Imaging Spectrometers (SIS) sensitive in the range 0.5$-$9
keV (140 eV resolution at E$=$6 keV), and two Gas Imaging Spectrometers
(GIS) with poorer energy resolution but with some efficiency up to 11
keV. The SIS observations were performed in 2-CCD Faint mode for
High bit rate and Bright mode for Medium bit rate
observations. The Faint mode data are converted into  Bright mode
data on the ground in order to combine them with the on-board Bright 
mode data. Thus all data analysis is performed on Bright data.
We used the eight independent data-sets
to extract the data in the regions of interest and then 
combined the datasets in 2 groups, one for all the GIS
and one for all the SIS, for the spectral analysis. 

Data preparation was done using the software package XSELECT
in FTOOLS (version v4.1) which allows selection of 
valid time intervals and removal of hot and flickering pixels. 
We used the newly reprocessed data of \rx ~as suggested by the ASCA 
processing team. The reprocessing was necessary since incorrect
orbit files were used in the first processing. The orbit files contain 
a time history of ASCA's orbital parameters and the use of out-of-date 
files could cause reduced signal to noise and inaccurate fluxes.
For the effective area and point spread function files we used
the latest versions available (xrt\_ea\_v2\_0.fits and 
xrt\_psf\_v2\_0.fits). Clean X-ray event lists were extracted 
using a magnetic cut-off 
rigidity threshold of 6 GeV c$^{-1}$ and the recommended minimum 
elevation angles and bright Earth angles to reject background 
contamination (Day et al. 1995). Events were extracted from 
within circles of radii 2.5 arcmin (SISs) and 6.25 arcmin (GISs). We
rejected about 98\% of cosmic ray events by using only SIS chip data
grades 0, 2, 3 and 4. Light curves for each instrument were visually 
inspected to exclude, if present, intervals with high background or 
data dropouts.  The individual SIS response matrices were generated
using ``sisrmg'' (v 1.1, April 97 version) which takes into account
the large time-dependence of various aspects of the SIS instruments 
responses. The SIS data were rebinned in the standard way to 512 channels 
applying Bright2linear with the lowest 17 channels ($<$ 0.6 keV) flagged
as bad. To convert SIS PHA (Pulse Height Analyzer) values to PI (Pulse 
Invariants) bins the file sisph2pi\_110397.fits was used.
The data were then regrouped so that no energy bin had fewer 
than 20 counts. Local background estimates were taken rather than the
deep background images supplied by the ASCA Guest Observer Facility 
(GOF). Even if the blank sky background subtraction gives
a better statistical quality, in the local background subtraction
both the particle and the galactic diffuse background are properly 
accounted for.
The background spectra of \rx ~were extracted with the same filters 
and parameters used to extract the source data. Background counts
were extracted from within annuli centered on the source and with
an inner radius of 6.25 (2.5) arcmin and outer radius of 11.25 (4) 
arcmin for the GISs (SISs). The details for both observations are 
given in Table 2. 

Spectral analysis was performed using XSPEC (v10.0) from the software 
package XANADU (\cite{arn96}). As mentioned earlier, given the low
statistics of our eight datasets (4 data-sets for each observation)
we fitted the spectral data, together with their respective response
files, after grouping them into 2 groups (one for the four GISs and one 
for the four SISs) in order to allow for different values of the
parameters fitting the GIS and SIS data.  The GIS data were fitted in the
0.8$-$9 keV band and the SIS data in the 0.6$-$8 keV band.
The spatial resolution of ASCA along with the low  signal-to-noise of 
the data permitted only a global single-temperature (\cite{ray77}) 
model to be fitted to the data. For the GIS dataset we fixed the 
hydrogen column density to the Galactic value  along the line of 
sight at the cluster position, N$_{H}=3.7\times10^{20}$
cm$^{-2}$ (\cite{dic90}) while it was left free for the SIS dataset.
The fitting procedure gave an anomalous measurement of
an order of magnitude higher value for the N$_{H}$, which may reflect
problems still present in the SIS calibration (see the web page
http://heasarc.gsfc.nasa.gov/docs/asca/cal\_probs.html).
The normalizations for the four GIS (SIS) spectra were constrained to be 
the same. The parameters varied to provide a fit to 8 spectra were 
therefore the 2 independent normalizations, the HI value, a single 
emission-weighted temperature, and the iron abundance. A best-fit 
cluster rest frame temperature of $kT=5.66^{+1.37}_{-0.58}$ keV
was measured  for the cluster gas and a value 
A$=0.43^{+0.25}_{-0.21}$ for the 
metallicity. All the errors are at the 68\% confidence levels
(1 $\sigma$), and all the fits have an acceptable reduced 
$\chi^{2}$=1.05$-$1.09. Figure~\ref{chisquare} shows the 
two-parameter $\chi^{2}$  contours for the cluster fractional
metallicity  and temperature. The derived unabsorbed
flux in the ASCA band is 
f$_{2-10keV}$$=$(1.78$\pm$0.11)$\times$10$^{-13}$ erg cm$^{-2}$ s$^{-1}$
corresponding to a luminosity in the cluster rest frame of
L$_{2-10keV}=$(8.19$\pm$0.43)$\times10^{44}$ h$_{50}^{-2}$ erg
s$^{-1}$, and to a bolometric luminosity
L$_{bol}$=(17.40$\pm$0.91)$\times10^{44}$ h$_{50}^{-2}$ erg
s$^{-1}$.
Converting to the 0.5$-$2.0 keV ROSAT band for comparison  with the ROSAT
HRI data and  with the ROSAT All-Sky survey data, and using the same 
assumptions  as above, we derive an unabsorbed flux in 0.5$-$2.0 keV equal to
f$_{0.5-2keV}=$(1.66$\pm$0.09)$\times10^{-13}$ erg cm$^{-2}$ s$^{-1}$,
and a luminosity in the cluster rest frame 
L$_{0.5-2keV}=$(4.57$\pm$0.24)$\times10^{44}$
h$_{50}^{-2}$ erg s$^{-1}$. This is a little higher than 
the flux reported in Henry (1997) of (1.13$\pm$0.31)
$\times$10$^{-13}$  erg cm$^{-2}$ s$^{-1}$
obtained from the RASS PSPC data.

\subsection{ROSAT Imaging}

X-ray observations of \rx ~were obtained with the ROSAT High Resolution
Imager (HRI, \cite{tru83}) in a single pointing in May 1997,
for a net live time of 170,853s. The HRI operates in the
ROSAT 0.1$-$2.4 keV energy band and provides an angular resolution of
$\sim$~4\arcsec\ (FWHM). 

We have extracted the total counts from the cluster in a circle 
of 125\arcsec\ radius ($\sim$1.03 h$_{50}^{-1}$ Mpc at the cluster 
redshift) centered at the 
position  $\alpha$(J2000)~=~17$^{h}$16$^{m}$48\fs2 and
$\delta$(J2000)~=~67\deg08\arcmin22\farcs4. This position is the
centroid of the X-ray emission as determined by
the maximum likelihood method used in the IRAF/PROS (v 2.5.p1) 
task {\it xplot.xexamine}. The position of the X-ray centroid
is fully consistent with the optical position of the cD galaxy (\#20
in Table 1) within the uncertainties of the HRI. The displacement in
right ascension is 1.7\arcsec to the west and in declination is
0.4\arcsec to the north of the brightest cluster galaxy.
Background counts were taken from three different regions free 
of sources within circles of 240\arcsec\ radius. There are 788 $\pm$ 117 
net counts in 0.1$-$2.4 keV in the cluster region. Assuming for
the temperature and the iron abundance of the gas the values 
kT$=$5.7 keV and A$=$0.4, as determined from ASCA data,
and a hydrogen column density along the line of sight 
N$_{H}$=3.7$\times$10$^{20}$ cm$^{-2}$, we determine an 
unabsorbed flux in 0.5$-$2.0 keV of
f$_{0.5-2.0keV}=$(1.42$\pm$0.21) $\times$ 10$^{-13}$ erg
cm$^{-2}$ s$^{-1}$. The X-ray luminosity in the 0.5$-$2.0 keV
cluster rest frame is
L$_{0.5-2keV}=$(3.94$\pm$0.58)$\times$10$^{44}$ h$_{50}^{-2}$
erg s$^{-1}$. Both the flux and luminosity are in excellent agreement
within the uncertainties with the ASCA determined values (see Table 3
for a summary of fluxes and luminosities in the various energy bands). 

The high on-axis resolution of the HRI allows us to study in more
detail the morphology of the X-ray emission from \rx. To
take full advantage of the intrinsic resolution of the instrument we
created an HRI image with $2.5\times 2.5$ arcsec$^2$ pixels using
software kindly provided by Steve Snowden. In addition to filtering
the raw event list, the code also produces maps of the expected HRI
particle background and the effective exposure time.
In Figure~\ref{xcontours} we show the X-ray iso-intensity contours from the HRI
observation overlaid on an optical CCD image of the cluster. The 
HRI counts image was smoothed with a Gaussian filter of 5 arcsec
width ($1\sigma$) and then corrected for particle background events and
exposure time variations.

As shown in Figure~\ref{xcontours}, the HRI reveals that \rx ~is
morphologically complex at X-ray wavelengths too. The emission is clearly
extended over a scale of roughly 1 arcmin ($\approx 500$ h$^{-1}_{50}$
kpc), and its shape is indicative of the non-regularity of this cluster. 
While the strongest emission originates from an almost circular peak
whose location is consistent with the position of the cD, the lower flux
contours show a pronounced centroid shift and become significantly
elongated in the same NE-SW direction as the distribution of the
galaxies\footnote{The faint source about 50 arcsec to the SE of the cluster
centre is most likely not related to the cluster.}. The elongated shape
of the emission coincides qualitatively with the dark matter contours
determined by Clowe et al. (1998), shown in the bottom panel of 
Figure~\ref{xcontours}.

\subsection{\bf X-ray Profile}

Even if $\beta$ models applied to this cluster are poorly constrained,
we will attempt in the
following  to extract an X-ray profile and derive an X-ray mass to 
compare to the weak lensing mass measured  by Clowe et al. (1998).
A radial profile of the X-ray emission is obtained by summing the HRI
counts in concentric annuli of variable width centered on the peak of
the X-ray emission and dividing by the area of the annuli.  This is
possible out to a radius of 125\arcsec\  ($\sim$1.03 h$_{50}^{-1}$ Mpc 
at the cluster redshift) at
which point the profile becomes indistinguishable, within the
noise, from the background level. The profile is then fit with a
$\beta$ model (\cite{cav76}) described by
the standard form $S(r) = S_0(1+r^2/r_c^2)^{-3\beta+1/2}$, with
S$_{0}$~= central surface brightness, $r_{c}$ = core radius
and $\beta$~=~slope parameter. The values of the best fit (90\%
confidence level) are S$_{0}$~= 3.24 $\times 10^{-13}$ erg cm$^{-2}$ 
s$^{-1}$ arcmin$^{-2}$, $r_c = 6\farcs8\pm6\farcs1$
and  $\beta$ = 0.42$^{+0.14}_{-0.09}$. Both the core radius and the
$\beta$ parameter are poorly constrained given the irregular
X-ray morphology.  At the distance of the cluster this core radius 
corresponds to 56 h$_{50}^{-1}$ kpc with a similar size error
(from 5.7 to 106 kpc). The size of the core radius could indicate 
the presence of a cooling flow. However, there is no evidence for 
excess emission in the very core of the cluster, thus any cooling 
flow must be of marginal significance. The existence of a cooling
flow in \rx ~would also be inconsistent with the emerging view that 
cooling flows form preferably in dynamically undisturbed clusters
(see Peres \etal among others). A fit of a point source plus 
the $\beta$ model to the data results in a contribution of less
than 15\% of the point-like component. To confirm or disprove the
presence of a central excess a spectral-spatial fit would be required.
This is not possible with the X-ray data in hand,  we have nonetheless 
attempted to fit a 2-temperature model to the ASCA spectra. The addition
of the second temperature model does not improve significantly 
the fit with respect to the 1-Temperature model (F-test
probability  less than 2\%).

Small core radii (less than 40--50 kpc) and small $\beta$ values 
($\sim$ 0.5), such as this cluster seems to possess, have been observed 
in other clusters; see for instance MS1512$+$3647 
($r_c = 6\farcs9\pm1\farcs1$ (42.6$\pm$6.7 h$_{50}^{-1}$kpc), 
$\beta = 0.524\pm0.031$; \cite{han97})  or MS0440$+$0204 
($r_c = 6\farcs4\pm1\farcs1$ (26.7$\pm$4.5 h$_{50}^{-1}$ kpc), 
$\beta = 0.45\pm0.03$; \cite{gio98}). The small size of the core radii and  
$\beta$s for these X-ray selected clusters is an issue that will 
be investigated in future work. Here we use these two correlated
parameters to derive an estimate of the cluster mass.

\subsection{Mass estimate: X-ray}

The mass determined using X-ray data does depend on assumptions
involving spherical symmetry and hydrostatic equilibrium, 
thus some of the assumptions may not apply in this case. 
These assumptions are found to be valid 
on the average in N-body simulation studies of cluster formation 
by Schindler (1996) and Evrard et al. (1996). Moreover, from 
extensive simulations Evrard et al. (1996)
have shown that errors on cluster masses, using
scaling relations between mass and temperature, are typically of the
order of 20\% even in cases, like \rx,  where some substructure is
present. \cite{all96} caution  against the use of single-phase analyses
with ROSAT data, since the presence of a cooling flow with distributed
mass deposition implies that the central intracluster medium (ICM)
has a range of temperatures and densities at any particular radius:
i.e. the ICM is multi-phase. There is no evidence of any significant
cooling flow in \rx, thus we feel that the assumptions adopted for
the deprojection outlined below are largely correct. 
We should bear in mind though that the derived quantities using the
HRI profile should be interpreted with some caution given the
asymmetry of the X-ray contours map that suggest that \rx ~is not a
well relaxed cluster.

From the HRI surface brightness profile,  assuming a
constant temperature, we have derived the three-dimensional 
density distribution of the gas from the two-dimensional image.
The integrated mass in
all forms can then be derived as a function of radius directly from
the equation of the hydrostatic equilibrium
\begin{equation}
M(r) = - {r  k T_g \over G \mu m_p} \left[ {d\ln \rho_g \over d\ln r}
+ {d\ln T_g \over d\ln r}\right],
\label{mr}
\end{equation}
\noindent 
The radial dependence of
the gas density $\rho_g$ is given by the ROSAT HRI observations. $T_g$ 
is the intracluster temperature, $\mu$ is the mean molecular weight of
the gas, and $m_p$ is the proton mass. A constant intracluster
gas temperature of  5.66 keV as measured by ASCA is assumed.
Our estimates for the gas and gravitational mass within 125\arcsec\
($\sim$1.03 h$_{50}^{-1}$ Mpc at the cluster redshift) are
(8.9$\pm2.1)\times10^{13}$ h$_{50}^{-5/2} M_{\sun}$ and
(2.8$\pm0.3)\times10^{14}$ h$_{50}^{-1} M_{\sun}$. 
The corresponding gas mass fraction,
M$_{gas}$/M$_{tot}$=(31$\pm$8)\% h$_{50}^{-3/2}$, is 
higher than found by other investigators 
(11$-$25\% h$_{50}^{-3/2}$ \cite{dav95} and \cite{whi95}). 
More recently Evrard (1997) has published a revised
average value of 17\% for the gas fraction using the above referenced
compilations. There is not yet a consensus on the value of the hot gas
to total mass ratio in clusters, and on the fact whether this
ratio is  really representative of the fraction of baryons in the 
universe. The measurements seem to indicate that the gas fraction may 
vary from cluster to cluster and may even vary in the same cluster as 
a function of radius (see Donahue 1998). 

We compare now the total mass from X-rays to the mass derived 
from the gravitational shear signal by Clowe et al. (1998). These authors 
find in roughly the same region (their 500 h$^{-1}$kpc aperture
corresponds to 1 Mpc in our cosmology) a value of 
(5.2$\pm$1.8)$\times10^{14}$ h$_{50}^{-1} M_{\sun}$.
The weak lensing mass is computed assuming the
background galaxies lie on a sheet at z$=$2 and is higher, even
within the uncertainties, than the X-ray derived mass. We note however
that the Clowe et al. (1998)  mass determination takes into account 
also the second clump of matter associated with the NE group of galaxies 
which is clearly separated in their mass distribution map from the main 
cluster mass (see  Figure~\ref{xcontours}). The X-ray  gravitational 
mass comes only from the main optical clump centered on the cD. In addition
the weak lensing center of mass of the main cluster is displaced with
respect to the center of the light and X-ray peaks of the cluster;
it is located 27\arcsec east and 11\arcsec north of the cD (see
Figure~\ref{xcontours}) but consistent within the uncertainties
of the weak lensing positions. The weak lensing mass distribution 
also resembles a filament of structure elongated in the same direction 
as the chain of optical galaxies. 

\section{Results and Discussion}

In the following discussion we will focus on the
implications of measured quantities such as the high velocity 
dispersion, or low temperature, rather than derived quantities
such as the gravitational mass from the X-ray data.
First we would like to note that using three completely different
instruments (ROSAT PSPC, ASCA and ROSAT HRI) we obtain
within the uncertainties a consistent flux for the cluster.
This is more remarkable than it seems since one has
to keep in mind that the discovery data of \rx ~consisted of only 33 net
counts. 

There are not many X-ray selected clusters known at this high redshift; the
other published examples are the 2 clusters from the EMSS (\cite{gio90}),
MS1054$-$03 at 0.83 and MS1137$+$66 at 0.78 (\cite{gio94}).  
Donahue et al. (1998) have made a follow-up study of MS1054$-$03
similar to the one presented here. We will compare in what follows the
similarities and differences between \rx ~and MS1054$-$03. A more 
detailed study of the other Medium Survey cluster,
MS1137$+$66, has been submitted (Donahue et al. 1999).

\rx ~and MS1054$-$03, both at a redshift higher than 0.8, show 
a filamentary  appearance in their optical morphology.
The Medium Survey cluster has an optical elongation 
in the east-west direction with the HRI contours  extended over 
a region of about 1\arcmin ~in the same direction as the optical
galaxies (\cite{don98a}).  \rx ~also has an extended X-ray emission 
of about 1\arcmin, with the center of the X-rays on the cD galaxy 
and the elongation in the same direction as the optical galaxies. 

Similarly to  MS1054$-$03, \rx~ has  a high velocity dispersion 
which can be interpreted as a signature of non-virialization. However,
differently from MS1054$-$03, we are dealing here with a rather cool 
cluster  (kT$=$5.66 keV  against kT$=$12.3 keV).
The temperature of \rx, 5.66$^{+1.37}_{-0.58}$,  is commensurate 
with the predictions from its X-ray luminosity from the
L$_{X}-$T$_{X}$ relation by  David et al. (1993) and Arnaud and Evrard
(1998). 
There is an  extensive literature on the correlation between these two
basic and measurable quantities (Edge \& Stewart 1991, Ebeling 1993,
David et al. 1993, Fabian et al. 1994, Mushotzky \& Scharf 1997, 
Markevitch 1998, Arnaud and Evrard, 1998). 
Comparing the bolometric luminosity of \rx ~with the best
fit relation, log(L$_{X}$)$=$(2.88$\pm$0.15)
log(T/6keV) $+$(45.06$\pm$0.03) obtained by \cite{ae98},
analyzing a sample of 24
low-z clusters with accurate temperature measurements and absence of
strong cooling flows, we would expect for \rx ~a temperature of
6.9 keV. We measure 5.66$^{+1.38}_{-0.58}$. The L$_{X}-$T$_{X}$ 
relation does not  seem to evolve much with redshift since z$=$0.4 
(Mushotzky \& Scharf 1997; Henry 1997) and possibly since 
z$=$0.8 (see study on MS1054$-$03 by \cite{don98a}).
Modest evolution is predicted in some evolution
models with $\Omega_{o}=$1 if the gas is pre-heated before falling into
the cluster potential (Evrard and Henry, 1991; Kaiser, 1991).

Next we check the $\sigma-$T$_{X}$ relationship. The
measured velocity dispersion of \rx, 1522$^{+215}_{-150}$ km s$^{-1}$,
is much higher than expected from its temperature. 
A large number of authors (see Table 5 in Girardi et al., 1996, or
Table 2 in Wu, Fang and Xu, 1998, for an exhaustive list of
papers on the subject) have attempted to determine the 
$\sigma-$T$_{X}$  using different cluster samples
in order to test the dynamical properties 
of clusters. Using the derived value from the HRI data of 
$\beta$$=$0.42 and the measured temperature of 5.7 keV 
from the ASCA data for the temperature of the gas in the equation
$\beta = { \mu m_p \sigma_v^{2} \over k T_{gas} }$
we obtain a velocity dispersion of about 600 km s$^{-1}$.  Even
assuming energy equipartition between the galaxies and the gas in the
cluster ($\beta$$=$1) we still obtain a velocity
dispersion much lower than measured ($\sim$ 929$^{+107}_{-48}$ km s$^{-1}$ 
vs 1522 km s$^{-1}$). In other words in an isothermal potential 
the mass distribution of the cluster does not correspond to the 
isotropic one-dimensional velocity dispersion. This is rather at odds
with results of hot, high-z clusters recently observed. The velocity
dispersions of three high ($>$0.5) redshift clusters, MS1054$-$03, 
MS0016$+$16 and MS0451$-$03 (see discussion in Donahue et
al. 1998) all reflect their temperatures according to the above
equation with $\beta$$=$1.
Girardi et al. (1996) have derived a best fit
relation between the velocity dispersion and the X-ray temperature,
with more than 30\% reduced scatter with respect to previous work
(Edge and Stewart 1991; Lubin and Bahcall 1993; Bird, Mushotzky and
Metzler, 1995; Wu, Fang and Xu 1998, among others).
Girardi et al. (1996) have taken into account distortions in the velocity 
fields, asphericity of the cluster or presence of substructures to 
derive their best fit relation,
log($\sigma$)$=$(2.53$\pm$0.04)+(0.61$\pm$0.05)log(T). If we plug in
the temperature of \rx ~in the above relation, the resulting 
velocity dispersion value is discrepant with what is measured.
A temperature  of 11.5$^{+3.0}_{-1.6}$ keV  would be expected 
for $\sigma$=1522 km s$^{-1}$. 

\rx ~may be an example of cluster which has not reached virial
equilibrium, its dynamical state may be in large part dominated by
infall or merging and consequently the velocity dispersion is not
representative of the virial temperature of the cluster. If \rx ~is
actually composed of two or more distinct gravitational components
the situation should be apparent from the velocity histogram
(which is Gaussian from Fig. 2), unless
the separation of the cluster components is so small to be
statistically of little significance. The high value of the
velocity dispersion can be due to a number of causes.
If there is any fraction of infalling galaxies which are bound to the
cluster but not yet virialized, they could inflate the velocity
dispersion. These galaxies in the cluster would be moving
on radial orbits.
To see if any substructure is present in the redshift space we have 
considered the galaxies in Figure~\ref{image} lying outside 
a circle (0.5 Mpc radius)
centered on the cD, to the north-east and to the south-west
of a dividing line passing through the cD  with position
angle of about $+$135\deg ~East of North.
There are eleven galaxies to the NE (labeled in Table 1 as
1, 2, 3, 4, 5, 6, 7, 8, 9, 10 and 25) and nine galaxies to the SW
(labeled 27, 28, 29, 30, 31, 32, 33, 34 and 37). 
The mean redshift of the galaxies to the NE is 0.8107$\pm$0.0018 
(1 $\sigma$ on the mean), and the mean redshift of the galaxies 
to the SW is 0.8021$\pm$0.0027, which implies a difference 
in redshift between the two components of
$\Delta$z$=$0.0086$\pm$0.0032 (1425 km s$^{-1}$ in velocity space),
a small effect that is significant at 2.65 $\sigma$. It might well
be that  galaxies are infalling towards the
center of the cluster, and while some galaxies have already reached
(or crossed) the core region, some others are still moving along 
radial orbits towards its center.

We recall here that numerical simulations  have shown that
the universe is composed of a web of filaments and voids on the scale
of hundreds of megaparsecs. The clusters of galaxies form at the
intersections of these filaments and grow with time
from quasi-spherical systems to the spherical objects that
we observe today as gas, galaxies and dark matter potential wells.
Filamentary structures have been observed in less distant clusters 
(i.e. A2125, Wang et al. 1998; the ABCG 85/87/89 complex, 
Durret et al. 1998). The initial formation of protoclusters
is often described as matter flowing along filaments (\cite{bon96})
with the X-ray flux maximum at the impact point of two colliding streams
of matter. In this scenario the hot gas and the galaxies would not be in
hydrostatic equilibrium and the transient velocity dispersion could 
be higher than expected from a virialized system. We are probably 
witnessing this process in this very distant X-ray selected cluster,
and \rx ~might be more properly called a protocluster.

\acknowledgments

We are grateful to Drew Phillis and collaborators for the use
of his UCSCLRIS package and to Timothy Beers for the use of his 
ROSTAT program to calculate velocity dispersions. Steve Snowden kindly
provided us with the latest version of his {\sc CAST\_HRI} package
which we found very useful in the analysis of the data from 
our HRI observation. We thank an anonymous referee for helpful 
comments which improved this paper. Partial 
financial support from NSF grant AST95-00515, from NASA-STScI grants
GO-5402.01-93A  and GO-05987.02-94A, from NASA grants NAG5-1880 and 
NAG5-2523, from SAO contract SV4-64008 and from CNR-ASI grants
is gratefully acknowledged.

\pagestyle{empty}
\clearpage
\begin{figure}
\epsscale{.9}
\caption{The image is a 1024$\times$1024 subarray extracted from the
center of a 4500s exposure in the I-band taken by Luppino 
and Metzger with the UH 8K$\times$8K CCD mosaic-camera on the CFHT
prime focus. The exposure was made up of 5 dithered pointings
each 900s long. This image has a seeing of 0\arcsec.7 FWHM and spans 
7\arcmin.2$\times$7\arcmin.2 (3.5$\times$3.5 h$_{50}^{-1}$Mpc at the 
redshift of the cluster). North is up and East to the left.
\label{image}
}
\end{figure}

\begin{figure}
\epsscale{.9}
\figcaption{Distribution of velocities for the entire sample of 37
galaxies.  The curve shows a Gaussian distribution with width given 
by the derived $\sigma_{los}$ and normalized to the 37 galaxies 
considered cluster members. The K-S test plus a suite of other robust
diagnostics indicate the distribution is Gaussian with a probability
$>$90\%.
\label{vel_hist}
}
\end{figure}

\begin{figure}
\epsscale{.7}
\caption{Two-dimensional $\chi^{2}$ contours at 68.3\% and 90\% 
confidence levels ($\Delta\chi^{2}=$2.30 and  4.61) for the 
cluster iron abundance in units of solar abundance and temperature in keV. 
These contours represent the confidence contours of simultaneous fits to GIS
and SIS spectra binned such that each bin contains a minimum of 20 
counts. 
\label{chisquare}
}
\end{figure}

\begin{figure}
\caption{The upper panel shows a CFHT I-band image of the central region of 
  \rx ~with contours of the HRI flux overlaid. The HRI raw counts
  were smoothed with a Gaussian of 5 arcsec width ($1\sigma$) and then
  corrected for particle background contamination and exposure time
  variations. The remaining X-ray background in this image is
  $4.3\times 10^{-7}$ ct s$^{-1}$ arcsec$^{-2}$; the
  contours are placed at $1.8\times{\rm background}\times 1.15^n$.
  The bottom panel shows a Keck-II R-band image of the same area
  with the weak lensing contours from Clowe et al. (1998)
  overlaid. Either   panel covers an area of $2.75\times 2.75$
  arcmin$^2$ ($1.3\times 1.3$ h$^{-2}_{50}$ Mpc$^2$ at the 
  redshift of the cluster).
\label{xcontours}
}
\end{figure}

\clearpage
\tablecaption{Spectroscopic data for the member galaxies of \rx~}
\tablecaption{ASCA observing information }
\tablecaption{X-ray data quantities }

\clearpage
\begin{deluxetable}{rrrrrrl}
\footnotesize
\tablecaption{Spectroscopic data for the member galaxies of
\rx~\label{tbl-1}} 
\tablewidth{400pt}
\tablehead{
\colhead{ID} &  
\colhead{$\alpha$(J2000)} &
\colhead{$\delta$(J2000)} & 
\colhead{cz ({\it km/s})} & 
\colhead{cz ({\it err})} & 
\colhead{z} & 
\colhead{Spectral Features}
} 
\startdata
 1 &  17:17:19.5  & +67:11:48.9  &  241309  &  150  &  0.8049 & H$+$K, CN \nl
 2 &  17:17:14.7  & +67:11:35.3  &  244127  &  900  &  0.8143 & [OII] \nl
 3 &  17:17:14.2  & +67:11:39.7  &  241699  &   60  &  0.8062 & H$+$K, CN \nl
 4 &  17:17:09.7  & +67:10:48.9  &  244607  &  240  &  0.8159 & [OII] \nl
 5 &  17:16:59.1  & +67:10:18.6  &  246046  &  150  &  0.8207 & H$+$K, CN, H$\gamma$ \nl
 6 &  17:17:08.9  & +67:09:27.1  &  243857  &  270  &  0.8143 & H$+$K \nl
 7 &  17:17:01.0  & +67:09:32.1  &  244187  &  120  &  0.8145 & H$+$K, G-band \nl
 8 &  17:17:01.2  & +67:09:24.0  &  242418  &   90  &  0.8086 & H$+$K, CN \nl
 9 &  17:17:03.8  & +67:08:59.6  &  242118  &   90  &  0.8076 & [OII], H$+$K, H$\delta$ \nl
10 &  17:16:56.9  & +67:09:04.6  &  240110  &   90  &  0.8009 & H$+$K, CN \nl
11 &  17:16:55.1  & +67:09:05.1  &  240350  &   90  &  0.8017 & H$+$K, CN, H$\delta$ \nl
12 &  17:16:51.0  & +67:08:51.4  &  246016  &   90  &  0.8206 & H$+$K, CN, G-band \nl
13 &  17:16:54.5  & +67:08:42.5  &  239630  &   30  &  0.7993 & H$+$K, CN \nl
14 &  17:16:55.0  & +67:08:32.7  &  246315  &  900  &  0.8216 & [OII] \nl
15 &  17:16:51.4  & +67:08:27.6  &  241639  &  160  &  0.8060 & H$+$K, CN \nl
16 &  17:16:54.1  & +67:08:21.9  &  244817  &  900  &  0.8166 & CaII-break \nl
17 &  17:16:53.6  & +67:08:07.6  &  240199  &  900  &  0.8012 & H$+$K \nl
18 &  17:16:50.1  & +67:08:20.7  &  238191  &  180  &  0.7945 & H$+$K, CN \nl
19 &  17:16:50.1  & +67:08:22.4  &  239240  &  900  &  0.7980 & [OII] \nl
20 &  17:16:48.5  & +67:08:22.0  &  247515  &  450  &  0.8256 & cD GALAXY, H$+$K \nl
21 &  17:16:47.3  & +67:08:17.6  &  243467  &  180  &  0.8121 & H$+$K, CN, H$\delta$ \nl
22 &  17:16:49.1  & +67:08:22.0  &  246106  &  240  &  0.8209 & H$+$K, CN \nl
23 &  17:16:46.7  & +67:08:16.0  &  243078  &  120  &  0.8108 & H$+$K, CN \nl
24 &  17:16:44.4  & +67:08:11.0  &  247095  &  150  &  0.8242 & CaII-break, CN \nl
25 &  17:16:43.7  & +67:09:43.9  &  242838  &  900  &  0.8100 & [OII] \nl
26 &  17:16:39.0  & +67:08:33.0  &  242868  &   30  &  0.8101 & [OII], H$+$K, H$\delta$ \nl
27 &  17:16:36.9  & +67:08:28.9  &  238251  &   60  &  0.7947 &
[OII],H$+$K, CN, G-band, 4C 67.26 \nl
28 &  17:16:36.9  & +67:08:14.1  &  241039  &   90  &  0.8040 & [OII], CN, H$+$K, H$\gamma$ \nl
29 &  17:16:34.4  & +67:08:02.7  &  242118  &  900  &  0.8076 & [OII] \nl
30 &  17:16:23.5  & +67:07:43.4  &  242598  &  900  &  0.8092 & [OII] \nl
31 &  17:16:32.6  & +67:07:36.4  &  244157  &  150  &  0.8144 & [OII], H$+$K, G-band \nl
32 &  17:16:35.7  & +67:07:48.2  &  239630  &   90  &  0.7993 & [OII], CN, H$+$K \nl
33 &  17:16:37.6  & +67:07:30.4  &  241159  &  120  &  0.8044 & H$+$K \nl
34 &  17:16:42.6  & +67:06:42.9  &  237771  &  900  &  0.7931 & [OII] \nl
35 &  17:16:49.0  & +67:07:20.3  &  242928  &   60  &  0.8103 & H$+$K, CN, G-band, H$\gamma$ \nl
36 &  17:16:54.7  & +67:07:38.5  &  246945  &  300  &  0.8237 & H$+$K, CN, G-band \nl
37 &  17:16:58.5  & +67:06:52.2  &  237561  &   30  &  0.7924 & H$+$K \nl
\enddata
\end{deluxetable}

\clearpage
\begin{deluxetable}{cccc}
\footnotesize
\tablecaption{ASCA observing information \label{tbl-2}}
\tablehead{
\colhead{Instrument} & \colhead{Arcmin Radius} & \colhead{Cts 
s$^{-1}$} &  \colhead{Screened Exposure in s}}
\startdata
GIS2-obs1 & 6.25 & (3.618$\pm$0.750)$\times10^{-3}$ &  54,713 \nl
GIS2-obs2 & 6.25 & (5.471$\pm$1.154)$\times10^{-3}$ &  25,109 \nl
GIS3-obs1 & 6.25 & (7.604$\pm$0.824)$\times10^{-3}$ &  54,713 \nl
GIS3-obs2 & 6.25 & (8.781$\pm$1.214)$\times10^{-3}$ &  25,105 \nl
SIS0-obs1 &  2.5 & (5.171$\pm$0.596)$\times10^{-3}$ &  45,687 \nl
SIS0-obs2 &  2.5 & (6.278$\pm$0.865)$\times10^{-3}$ &  21,005 \nl
SIS1-obs1 &  2.5 & (4.691$\pm$0.619)$\times10^{-3}$ &  45,694 \nl
SIS1-obs2 &  2.5 & (5.393$\pm$0.905)$\times10^{-3}$ &  20,981 \nl
\enddata
\end{deluxetable}

\clearpage
\begin{deluxetable}{lccccc}
\footnotesize
\tablecaption{X-ray data quantities \label{tbl-3}}
\tablewidth{500pt}
\tablehead{
\colhead{Instrument}      & 
\colhead{f$_{0.5-2}$ keV  } &
\colhead{f$_{2-10}$  keV  } &  
\colhead{L$_{0.5-2}$ keV  } &
\colhead{L$_{2-10}$  keV  } &
\colhead{L$_{BOL}$   keV  } \\
\colhead{}                                    & 
\colhead{ erg cm$^{-2}$ s$^{-1}$} &
\colhead{ erg cm$^{-2}$ s$^{-1}$} &
\colhead{ erg s$^{-1}$}         &
\colhead{ erg s$^{-1}$}         &
\colhead{ erg s$^{-1}$}  
}

\startdata
PSPC & (1.13$\pm$0.31)10$^{-13}$ &   --          & (3.20$\pm$0.90)10$^{44}$ & --            &  -- \nl
ASCA & (1.66$\pm$0.10)10$^{-13}$ & (1.78$\pm$0.11)10$^{-13}$ & (4.57$\pm$0.24)10$^{44}$ & (8.19$\pm$0.43)10$^{44}$ & (17.4$\pm$0.91)10$^{44}$ \nl
HRI  & (1.42$\pm$0.21)10$^{-13}$ &   --          & (3.94$\pm$0.58)10$^{44}$ & --            & -- \nl
\enddata

\end{deluxetable}


\clearpage
\begin{thebibliography}{}
\bibitem[Allen et al.\ 1996]{all96} Allen, S.W., Fabian. A.C. and Kneib,
        J.P., 1996, MNRAS, 279, 615
\bibitem[Arnaud\  1988]{arn88} Arnaud K.A. 1988, in ``Cooling flows in
        Clusters and Galaxies'', Fabian, A.C. ed, Kluwer Academic Publ.
        Dordrecht, p. 31
\bibitem[Arnaud \ 1996]{arn96} Arnaud, K.A., 1966, Astronomical Data Analysis 
	Software and Systems V, eds. G. Jacoby and J.K. Barnes, ASP
        Conf. Ser., vol 101
\bibitem[Arnaud and Evrard \ 1998]{ae98} Arnaud, M. and Evrard, A.E., 
        submitted to MNRAS, astro-ph/9806353
\bibitem[Beers, Flynn and Gebhardt 1990]{bee90} Beers, T.C., Flynn,
        K., and Gebhardt, K., 1990, \aj, 100, 32
\bibitem[Bird, Mushotzky and Metzler \ 1995]{bir95} Bird, C.M.,
        Mushotzky, R.F. and Metzler, C.A., 1995, \apj, 453, 40
\bibitem[Bond, Kofman and Pogosyan, 1996]{bon96} Bond, J. R., Kofman,
        L. and Pogosyan, D., 1996, Nature, 380, 603
\bibitem[Bower et al. \ 1994]{bow94} Bower, R. G.,  Boehringer, H.,
        Briel, U., Ellis, R. S., Castander, F. J. and Couch, W. J.,
        1994, MNRAS, 268, 345
\bibitem[Bower et al. \ 1997]{bow97} Bower, R. G., Castander, F. J.,
        Couch, W. J., Ellis, R. S. and Boehringer, H., 1997, MNRAS, 
        291, 353
\bibitem[Cavaliere and Fusco-Femiano \ 1976]{cav76} Cavaliere, A. and 
        Fusco-Femiano, R., 1976,  A\&A, 49, 137
\bibitem[Clowe et al. \ 1998]{clo98} Clowe, D.I., Luppino, G.A.,
        Kaiser, N., Henry, J.P. and  Gioia, I.M., 1998, \apjl, 497, L61
\bibitem[Crawford and Fabian \ 1996]{cra96} Crawford, C.S. and Fabian,
        C.A., 1996, MNRAS, 282, 1483
\bibitem[Danese et al.\ 1980]{dan80} Danese, L., De Zotti, G. and
        di Tullio, G., 1980, A\&A, 82, 322
\bibitem[David et al. \ 1993]{dav93} David, L.P., Slyz, A., Jones, C.,
        Forman, W. and Vrtilek, S.D., 1993, \apj, 412, 479
\bibitem[David et al.\ 1995]{dav95} David, L.P., Jones, C. Forman, W.
        1995, \apj, 445, 578
\bibitem[Day et al. \ 1995]{day95} Day, C., Arnaud, K., Ebisawa, K., 
	Gotthelf, E., Ingham, J., Mukai, K. and White, N. 1995, ``The ABC
	Guide to ASCA Data Reduction, Fourth Version'', available by request 
	from  the ASCA GOF.
\bibitem[Deltorn et al. \ 1997]{del97} Deltorn, J-M, {
        Le\thinspace F\`evre~}, O.,
        Crampton, D. and Dickinson, M., 1997, \apj, 483, L21
\bibitem[Dickey and Lockman \ 1990]{dic90} Dickey, J.M. and Lockman,
        F.J., 1990, ARAA, 28, 215
\bibitem[Donahue \ 1998]{don98} Donahue, M., 1998, in ``Origins''
        ASP Conference Series, eds., C.E. Woodward, J.M. Shull, 
        and H. Thronson, vol. 148, pages 109-126
\bibitem[Donahue et al. \ 1998]{don98a} Donahue, M., Voit, G.M., Gioia,
        I., Luppino, G., Hughes, J.P., and Stocke, J.T., 1998, ApJ, 
        502, 550
\bibitem[Donahue et al. \ 1999]{don99} Donahue, M., Voit, G.M.,
	Scharf, C.A., Gioia, I.M., Mullis, C.R., Hughes, J.P., and 
	Stocke, J.T., 1999, ApJ, submitted
\bibitem[Durret et al. \ 1998]{dur98} Durret, F., Forman, W., Gerbal, D.,
	Jones, C. and Vikhlinin, A., 1998, A\&A, 335, 41 
\bibitem[Ebeling \ 1993]{ebe93} Ebeling, H., 1993, phD Thesis, MPE
\bibitem[Ebeling et al. \ 1998]{ebe98} Ebeling, H., Jones, L.R.,
        Perlman, E., Scharf, C., Horner, D., Wegner, G., Malkan, M. and
        Mullis, C.R., 1998, submitted to \apj
\bibitem[Edge and Stewart \ 1991]{edg91} Edge, A.C., and Stewart,
        G.C., 1991, MNRAS, 252, 414
\bibitem[Evrard and Henry \ 1991]{evr91} Evrard, A.E. and Henry, J.P.,
        1991, \apj, 383, 95
\bibitem[Evrard et al.\ 1996]{evr96} Evrard, A.E., Metzler, C.A. and
        Navarro, J.F. 1996, \apj, 469, 494
\bibitem[Evrard \ 1997]{evr97} Evrard, A.E., 1997, MNRAS, 292, 289
\bibitem[Fabian et al. \ 1994]{fab94} Fabian, A.C., Crawford, C.S.,
        Edge, A.C., Mushotzky, R.F., 1994, MNRAS, 267, 779
\bibitem[Frenk et al. \ 1990]{fre90} Frenk, C.S, White, S.D.M.,
        Efstahtiou, G. and Davis, M., 1990, ApJ, 351, 10
\bibitem[Gioia et al. \ 1990]{gio90} Gioia, I.M., Maccacaro, T.,
        Morris, S. L., Schild,  R.E., Stocke, J.T., Wolter, A. and 
        Henry, J.P., 1990, \apjs, 72, 567
\bibitem[Gioia and Luppino \ 1994]{gio94} Gioia, I.M. and Luppino,
        G.A., 1994, \apjs,  94, 583
\bibitem[Gioia et al. \ 1998]{gio98} Gioia, I.M., Shaya, E.,
        Le\thinspace F\`evre, O.,
        Falco, E.E., Luppino, G.A. and Hammer, F., 1998, \apj, 497, 573
\bibitem[Girardi et al. \ 1996]{gir96} Girardi, M., Fadda, D.,
        Giuricin, G., Mardirossian, F. and Mezzetti, M., 1996,
        \apj, 457, 61
\bibitem[Hamana et al.\ 1997]{han97} Hamana, T., Hattori, M., Ebeling,
        H., Henry, J.P., Futamase, T. and Y. Shioya, 1997, \apj, 484, 574
\bibitem[Henry et al. \ 1997]{hen97} Henry, J.P., Gioia, I.M., Mullis, C.R.,
	Clowe, D.I., Luppino, G.A., Boehringer, Briel, U.G., Voges, W. 
	and Huchra J.P. 1997, \aj, 114, 1293
\bibitem[Henry \ 1997]{hen97b} Henry, J.P., 1997, \apjl, 489, L1
\bibitem[Kaiser \ 1991]{kai91} Kaiser, N., 1991, \apj, 383, 104
\bibitem[Le\thinspace F\`evre et al. \ 1996]{olf96} Le\thinspace F\`evre, O., 
        Deltorn, J.M., Crampton, D. and Dickinson. M., 1996, \apj, 471, L11
\bibitem[Lubin and Bahcall \ 1993]{lub93} Lubin, L.M. and Bahcall,
        N.A., 1993, \apj, 415, L17
\bibitem[Lubin et al. \ 1998]{lub98} Lubin, L.M., Postman, M. and Oke,
        J.B., 1998, Ratnatunga, K.U., Gunn, J.E., Hoessel, J.G.,
	Schneider, D.P., 1998, \aj, 116, 584
\bibitem[Markevitch \ 1998]{mar98} Markevitch, M., 1998, \apj, 504, 27
\bibitem[Mushotzky and Scharf \ 1997]{mus97} Mushotzky, R.F. and Scharf,
        C.A., 1997, \apj, 482, L13        
\bibitem[Oke et al. \ 1995]{oke95} Oke, J.B., Cohen, J.G., Carr, M.,
        Cromer,  J., Dingizian, A., Harris, F.H., Labrecque, S., Lucinio, R., 
        Schaal, W., Epps, H., and Miller, J., 1995, PASP, 107, 375
\bibitem[Oke et al. \ 1998]{oke98}  Oke, J.B., Postman, M., Lubin, L.,
        1998, \aj, 116, 549
\bibitem[Peres et al. 1998]{per98} Peres, C.B., Fabian, A.C., Edge,
	A.C., Allen, S.W., Johnstone, R.M., and White, D.A., 1999,
	 MNRAS, in press
\bibitem[Postman et al. 1998]{pos98}  Postman, M., Lubin, L.M. and
        Oke, J.B., 1998, \aj, 116, 560
\bibitem[Raymond \& Smith \ 1977]{ray77} Raymond, J.C. and Smith,
        B.W., 1977 \apjs, 35, 419
\bibitem[Rosati et al. \ 1995]{ros95} Rosati, P., Della Ceca, R., 
        Norman, Colin and Giacconi, R., 1995, \apj, 445, L11
\bibitem[Rosati et al. \ 1998]{ros98} Rosati, P., Della Ceca, R.,
        Norman, Colin and Giacconi, R., 1998 \apj, 492, L21
\bibitem[Rosati et al. \ 1999]{ros99} Rosati, P.,  Stanford, S.A.,
	Eisenhardt, P.R., Elston, R., Spinrad, H., Stern, D. and Dey, A.,
	1999, AJ, submitted
\bibitem[Smail and Dickinson \ 1995]{sma95} Smail, I. and Dickinson,
        M., 1995, \apj, 455, L99
\bibitem[Schindler \ 1996]{sch96} Schindler, S. 1996, A\&A, 756, 305
\bibitem[Stanford et al. \ 1997]{sta97} Stanford, S. A., Elston, R.,
        Eisenhardt, P. R., Spinrad,  H., Stern, D. and Dey, A., 1997, 
        \aj, 114, 2232
\bibitem[Tanaka et al. \ 1994]{tan94} Tanaka, Y., Inoue, H., and Holt,
	S.S. 1994, PASJ, 46, L37
\bibitem[Tr\"umper\ 1983]{tru83} Tr\"umper, J., 1983, Adv. Space. Res.
	2, 142
\bibitem[Van Haarlem et al. \ 1997]{har97} Van Haarlem, M.P., Frenk,
        C.S.  and White, S.D.M, 1997, MNRAS, 287, 817
\bibitem[Voges et al. \ 1996]{vog96} Voges, W., et al., 1996, MPE
        Rep, 263, 637
\bibitem[Wang et al. \ 1998]{wan98} Wang, Q.D., Connolly, A.J., and 
	Brunner, R.J. 1998,\apjl, 487, L13
\bibitem[White and Fabian \ 1995]{whi95} White, D.A. and Fabian, A.C.
        1995, MNRAS, 273, 72
\bibitem[White et al. \ 1997]{wjf97} White, D.A., Jones, C. \& Forman, W., 
	1997, MNRAS, 292, 419.
\bibitem[Wu et al. \ 1998]{wfx98} Wu, X-P, Fang, L-Z and Xu, W., 1998,
        A\&A, 338, 813

\end{thebibliography}
\end{document}